\documentclass[aps]{revtex4}

\begin{document}
\draft

%
%

\preprint{Nisho-07/2}
\title{Two Dimensional Quantum Well of Gluons\\
in Color Ferromagnetic Quark Matter}
\author{Aiichi Iwazaki}
\address{Department of International Politics and Economics, Nishogakusha University, Ohi Kashiwa Chiba
  277-8585,\ Japan.} 
\date{May. 7, 2007}
\begin{abstract}
We have recently pointed out that color magnetic field is generated in dense quark matter, i.e.
color ferromagnetic phase of quark matter. Using light cone quantization,
we show that gluons occupying the lowest Landau level under the color magnetic field
effectively form a two dimensional 
quantum well ( layer ), in which
infinitely many zero modes of the gluons are present.
We discuss that the zero modes of the gluons 
form a quantum Hall state by 
interacting 
repulsively with each other,
just as electrons do in semiconductors.
Such a ferromagnetic quark matter with the
layer structure of the gluons is a possible origin of extremely
strong magnetic field observed in magnetars.
\end{abstract}	
\hspace*{0.3cm}
\hspace*{1cm}

\maketitle
\section{introduction}
Dense quark matters have been extensively analyzed
and have been shown to possess various interesting phases.
Most of these phases arise from the dynamical effects of quarks,
e.g. the condensation of diquark pairs. The phases with the diquark condensation
are called as color superconducting phases\cite{cs}.
The phases have been argued to arise in sufficiently dense quark matters,
where perturbative approximations are reliable
owing to the asymptotic freedom of QCD.
( Non interacting free quark gas forms a Fermi surface, but
it is unstable against to an attractive force produced by
one gluon exchange between
a diquark channel, even if it is fairly weak. 
This instability leads to 
the condensation of the diquark pairs. Since the pairs carry color charges,
color superconducting states arise. )
On the other hand,
quark matters are composed of not only quarks but also gluons.
Our previous analyses\cite{cf} of the gluons in the dense quark matters
have revealed a color ferromagnetic phase in which
color magnetic field, $B$, is generated spontaneously.
The generation happens by not alignment of quark spins\cite{tatsumi}, but
the vacuum fluctuation of gluons at one loop order;
an effective potential of color magnetic field obtained with one loop approximation
shows a non-trivial minimum\cite{savvidy} at $B\neq 0$.
The loop expansions
are also reliable in the dense quark matters. 
Thus, we must determine which phase arises, the color superconducting phase or the color
ferromagnetic phase in the sufficiently dense quark matter.
We have shown\cite{cf,cf2} that the color superconducting phase is favored in the limit of
infinitely dense quark matter.
We have also shown that the color ferromagnetic phase appears as an energetically more favorable
state than the color superconducting state 
as we decrease the baryon densities of the quark matters.
Recent analyses\cite{gapless} of the instability of the gapless color superconductors suggest the existence of the phase transition from
the color superconducting state to the color ferromagnetic state with decreasing the baryon density.

The quark matter may be present in the core of neutron stars.
For instance, neutron stars with mass $\simeq 1.6M_{\odot}$,
and radius $\simeq 10$km, have the average density
$\simeq 2.7 \rho_{n}$, where $\rho_{n}$ represents
the normal nuclear density approximately given by $2.8\times 10^{14}$g/cm$^3$.
If the density in the core of the neutron stars 
reaches the density $\simeq 6\rho_n$,
nucleons overlap and they would melt into quark matter.
Models\cite{n-star} of neutron stars involving only nuclear components
predict that the density at the center reaches ($6\sim 10$)$\rho_n$, or more.
Hence, it is very probable that the cores of neutron stars
involve quark matter. Such a quark matter would not be sufficiently dense
for the color superconducting states to appear.
But it would be sufficiently dense for the color ferromagnetic states to appear.
We have shown in the previous paper\cite{magnetar} that extremely strong electromagnetically 
magnetic fields ( $\sim 10^{15}$ Gauss ) observed in magnetars
can be produced by the color ferromagnetic quark matter. 
Thus, the detail examination of such a quark matter is an important issue.
In the present paper we show that the gluons form effectively a
two dimensional quantum well in the color ferromagnetic quark matter.
The presence of the quantum well is an important ingredient for the generation of
the strong magnetic field in the magnetars.

As was shown about 30 years ago,
the naive ferromagnetic state with color magnetic field is unstable\cite{savvidy,nielsen} in QCD.
Gluons occupying the lowest Landau level under the color magnetic field
have imaginary energies, that is, the gluons are unstable. The situation is very similar to the case of Higgs models
in which Higgs fields have imaginary energies in a naive false vacuum without 
any condensation of the fields. The real vacuum is a condensed state of the Higgs field.
Thus, we expect that such unstable gluons condense to form a stable ground state
under the color magnetic field.
But the problem of finding the stable ground state\cite{cf,ninomiya,06} is not so easy because
there are infinitely many unstable gluons characterized by their angular momenta, $m$ ( $\infty >m\geq 0$ )
in the lowest Landau level. All of them condense to form an appropriate ground state.  
In the Higgs model, only the spatially uniform component of the Higgs field condenses
to form a uniform ground state. Thus, it is easy to find the classical solution of the Higgs field
representing the ground state.
In this paper we show that the unstable gluons condense to form effectively a two dimensional
quantum well. Namely, they make a layer perpendicular to the color magnetic field.
Furthermore, we show that there are infinitely many excited states of gluons with zero energy ( zero modes )
in the quantum well. 
In other words, there exist infinitely many degenerate
ground states of the gluons. 
Each of the zero modes induces a color magnetic field
to screen partially the original one. Thus,
it apparently seems to indicate expelling the magnetic field or squeezing it
by exciting the zero modes just like superconductors. 
This is because the condensed state of the unstable gluons is expected to show a Meissner effect
just as in Higgs models. 
But, expelling or squeezing the magnetic field is energetically unfavorable
in the gauge theory.
We show that contrary to superconducting states,  
a quantum Hall state of the gluons comes out as a gapped stable state
in the quantum well.

In condense matter physics
two dimensional quantum wells are fabricated by connecting two semiconductors,
for example, GaAs and AlGaAs\cite{qhs}. The junction is a surface with small width ( $\sim 10^{-6}$cm ) in which
electrons are confined. Their motions in the direction perpendicular to
the surface ( longitudinal direction ) are not allowed 
as far as we are concerned with low energies, or low temperatures.
On the other hand, their motions in two dimensional directions parallel to the surface ( transverse directions )
are allowed. These behaviors result from the following feature of the excitation energies of electrons.
Since electrons are confined in the quantum well with small width in the longitudinal direction,
any excitation energies in the direction are finite and much larger than
ones we are concerned with. On the other hand, excitations energies in the two dimensional transverse directions
start to grow at zero energy, in other words, the excitations are gapless.
This is a typical feature of electrons confined in
the two dimensional quantum well fabricated in semiconductors. 

As we will show, the similar excitation spectra of the gluons to those of the electrons
are obtained 
as a result of the condensation of the unstable gluons under the color magnetic field.
This implies that the gluons form effectively a two dimensional quantum well in the ferromagnetic dense
quark matter.

In the next section, we introduce a light cone formulation\cite{lc,cone} 
of SU(2) gauge theory whose gauge fields are referred as gluons.
We extract "zero modes"\cite{yamawaki,yama} by using finite volume 
in the longitudinal direction and neglecting the zero modes\cite{thorn}.
This leads to a simple form of Hamiltonian with the use of
the light cone gauge, but we loose a merit of the light cone formulation;
the real vacuum is a Fock vacuum. We do not address 
the spontaneous generation of the color magnetic field
in the light cone formulation. 
We simply assume the presence
of the field.
In the section (3) we pick up only states of gluons
in the lowest Landau level. Only the states are relevant to
the formation of the ground states of the gluons.
In the section (4) we examine the classical 
and quantum structures of the ground states.
We find that the ground states are not coherent states, but
approximate eigenstates of number operators of the gluons. 
In the section (5) we analyze excitation energies in the
ground states. We find that two dimensional quantum well is formed
effectively. We also find that there are infinitely many states with zero energy in the well.
In the section (6) we examine the effects of such states on the structure of
the ground states.
Although each of the states partially screen the color magnetic field, 
their excitations do not completely screen the field but
form a fractional quantum Hall state in the well.
In the last section (7) we summarize
our results.

\section{light cone formulation of SU(2) gauge theory}
First of all, we explain our notations 
in the light cone quantization\cite{lc,cone} of gauge fields.
We use the light cone time coordinate, $x^+=(x^0+x^3)/\sqrt{2}$ and longitudinal
coordinate, $x^-=(x^0-x^3)/\sqrt{2}$. Transverse coordinates are denoted by $x^i$,
or $\vec{x}$.
We assume a finite length, $-L\leq x^- \leq L$,
in the longitudinal space and impose a periodic boundary condition such that
$A^a_j(x^{-}=L)=A^a_j(x^{-}=-L)$. Then,
corresponding momentum becomes discrete denoted by $p_n^+=n\pi/L$ with integer $n$.
Light cone components of gauge fields, $A^+,\,A^-$, $A_i$, are defined similarly.

Then, the Hamiltonian, $H$, with the light cone gauge, $A^+=0$, is given by

\begin{equation}
\label{H}
H = \frac{1}{4} F_{ij}^a\, F_{ij}^a + 
\frac{g^2}{2}\rho^a\, \frac{1}{(-\partial_-^2)}\,\rho^a ,
\end{equation}
with field strength, $F_{ij}^a=\partial_iA^a_j-\partial_jA^a_i+g\epsilon_{abc}A^{b}_iA^{c}_j$,	
where color indices $a$ run from $1$ to $3$ and space indices, $i,j$ run from $1$ to $2$.
Color charge, $\rho^a$, is defined by

\begin{equation}
\rho^a\equiv (D_iA_i)^a+\rho_{quark}^a=(\partial_i\delta^{ab}+g\epsilon^{a3b}A_i^B)A_i^b+\rho_{quark}^a
\end{equation}
where $\rho_{quark}$ denotes the contribution of quarks. 
Here we treat it classically and assume it being spatially uniform and pointing to $\sigma_3$ in color space.
$A_i^B$ denotes the gauge potential of color magnetic field
generated spontaneously, which is assumed 
to direct into $\sigma_3$ in color SU(2);
$B=\partial_1A_2^B-\partial_2A_1^B$. In this paper we do not address a question of the spontaneous generation of the 
magnetic field, $B$, in the light cone formulation. We simply assume the presence of $B$. 

In the above equations we have neglected a dynamical gauge potential, $A_i^{a=3}$ aside from the classical 
one, $A_i^B$ since it does not couple directly with $A_i^B$. We have only taken 
dynamical gauge fields, $A_i^{a=1,2}$ perpendicular in color space to the color magnetic field.
They form Landau levels under $B$.

We make a comment that our treatment of "zero mode"\cite{yamawaki,yama}
in the light cone quantization
is similar to the one used by Thorn\cite{thorn}: 
We quantize gauge fields 
in the finite volume, $-L\leq x^{-}\leq L$, and neglect "zero modes" of the fields.
Consequently, Hamiltonian becomes a simple form involving
at most quartic terms of creation or annihilation operators
in addition to quadratic ones. As has been shown in a two dimensional model
of scalar field\cite{thorn} and a Higgs model of complex scalar field\cite{lcq}, the true ground states can be gripped even if
we neglect the "zero modes" of the fields, at least in the limit of $L\to \infty$.  
We assume that it also holds in the gauge theory.
We may justify 
neglecting the "zero modes" in the analysis of dense quark matter as follows.
That is, our concern is not the real vacuum, but a ground state of gluons
in dense quark matter. 
The "zero modes" may play an important role in the real vacuum of strongly interacting gluons.
But, they may not play such a role 
in a ground state of gluons weakly interacting with each other in the dense quark matter. 
Typical energy scale of quarks and gluons in the quark matter 
is given by the chemical potential of the quarks and is much larger than $\Lambda_{QCD}$.
In such dense quark matter the "zero modes" do not play an important role for
realizing the ground state.
It is similar to the case of QCD at high energy scattering
where the typical energy scale is much higher than $\Lambda_{QCD}$. Hence,
the "zero modes" do not play an important role 
for realizing so-called color glass condensate\cite{cgc}.
Therefore, it is reasonable to neglect the "zero modes" 
of the gluons in the quark matter.

In the light cone gauge only dynamical variables are transverse components, $A_i^a$ of gauge fields.
This can be expressed in terms of creation and annihilation operators,

\begin{equation}
A_i^b=\sqrt{\frac{\pi}{L}}\sum_{p^+>0}\frac{1}{\sqrt{2\pi p^+}}
(a_{i,p^+}^b(x^+,\vec{x})e^{-ip^+x^-}+ a_{i,p^+}^{b\dagger}(x^+,\vec{x})
e^{ip^+x^-}),
\end{equation}
with $p^+ =\pi n/L$ ( integer, $n\geq 1$ ),
where operators, $a_{i,p^+}^l$, satisfy the commutation relations,
$[a_{i,p^+}^b(x^+,\vec{x}),
a_{j,k^+}^{\dagger c}(x^+,\vec{y})]=\delta_{ij}\delta^{bc}\delta_{p^+,k^+} \delta(\vec{x}-\vec{y})$,
with other commutation relations being trivial. 
As we have mentioned,
we have neglected the "zero modes", $p^+=0$, of the gauge fields. 
  
Then, the gauge fields satisfy the equal time, $x^+$, commutation relation,

\begin{equation}
[\partial_-A_i^a(x^+,x^-,\vec{x}),A_j^b(x^+,y^-,\vec{y})]=
-i\delta_{ij}\delta^{ab}\delta(\vec{x}-\vec{y})\Bigl(\delta(x^- -y^-)-\frac{1}{2L}\Bigr),
\end{equation}
where the last factor, $1/2L$, in the right hand side of the equation 
comes from neglecting the "zero modes"
of the gauge fields.

We should mention that the second term in $H$ represent a Coulomb interaction. 
It is derived by solving a constraint equation, $\partial_-^2 A^{-,a}=\rho^a$,
that is, Gauss law associated with the light cone gauge condition, $A^+=0$.
In order to assure that the gauge field, $A^-$ is periodic in $x^-$,
the zero mode of $\rho$ ( $\rho\propto \sum_{n=\mbox{integer}}\rho_n\, e^{i\pi nx^-/L}$ )
must vanish; $\rho_{n=0}=0$. Then, the operation of $1/(-\partial_-^2)$ is 
well defined. The condition of $\rho_{n=0}=0$ implies that the color charge, 
$\int_{-L}^{L} dx^- \rho$,
vanishes and is consistent with our postulate, $A^-_{n=0}=0$. 

We now rewrite the Hamiltonian in terms of "charged vector fields", $\Phi_i=(A_i^1+iA_i^2)/\sqrt{2}$,
which are decomposed into spin parallel ( $\Phi_p=(\Phi_1+\Phi_2)/\sqrt{2}$ )
and anti-parallel components ( $\Phi_{ap}=( \Phi_1-\Phi_2)/\sqrt{2}$ ).
These fields transform as Abelian charged fields under the $U(1)$ gauge transformation, 
$A_i\to U^{\dagger}A_iU+U^{\dagger}\partial_iU$ with $U=exp(i\theta \sigma_3)$.

Then, using the fields, $\Phi_{p}$ ( $\Phi_{ap}$ ), we obtain the following Hamiltonian,

\begin{eqnarray}
\label{H'}
H=\frac{1}{2}B^2&+&\Phi_{p}^{\dagger}(-\vec{D}^2-2gB)\Phi_{p} +\Phi_{ap}^{\dagger}(-\vec{D}^2+2gB)\Phi_{ap} \\ \nonumber
&+&\frac{g^2}{2}(|\Phi_{p}|^2-|\Phi_{ap}|^2)^2+\frac{g^2}{2}\rho\,\frac{1}{(-\partial_-^2)}\,\rho\,\,\,, 
\end{eqnarray}
with $\vec{D}=\vec{\partial}+ig\vec{A^B}$,
where $\rho$ is given by

\begin{equation}
\rho=i(\Phi_{p}^{\dagger}\partial_-\Phi_{p} -\partial_-\Phi_{p}^{\dagger}\Phi_{p} +
\Phi_{ap}^{\dagger}\partial_-\Phi_{ap} -\partial_-\Phi_{ap}^{\dagger}\Phi_{ap})+\rho_{quark}\,\,.
\end{equation}

The first term in eq(\ref{H'}) represents the classical energy of the color magnetic field, and second ( third ) term
does the kinetic energy of the charged gluons with spin parallel ( anti-parallel ) under the color
magnetic field, $B$. The terms, $\pm 2gB|\Phi_{p,ap}|^2$, represent anomalous magnetic moments of the charged gluons\cite{nielsen}.
The forth term represents the energy of the repulsive self-interactions. The last term
represents the Coulomb energy coming from the second term with $\rho^{a=3}$ component in eq(\ref{H}).

\section{gluons in the lowest Landau level}

As we have mentioned, the ground state of the gluons in the system is determined by
the gluons occupying the lowest Landau level.
Since eigenstates of the operator $\vec{D}^2$ are classified by Landau levels,
the second and the third terms in the Hamiltonian of eq(\ref{H'}) can be rewritten as,

\begin{equation}
\sum_{n=0,1,2,,,}\Big(\Phi_{p,\,n}^{\dagger}(2n-1)gB\Phi_{p,\,n}+\Phi_{ap,\,n}^{\dagger}(2n+3)gB\Phi_{ap,\,n}\Big),
\end{equation}
where the fields, $\Phi_{p,\,n}$ ( $\Phi_{ap,\,n}$ ) denote operators in the Landau level specified by integer $n$.
( We have implicitly assumed integration over the transverse directions in the above equation. )

Now, we take only the field, $\Phi_{p,\,n=0}$ in the lowest Landau level, $n=0$,
that is, the component having negative kinetic energy. 
Obviously, the other components of the gluons have positive energies so that
the ground state is the state with no such gluons.
The gluons with negative kinetic energies play important roles for the formation of the ground state,
at least, in the limit of strong magnetic field, $B$.
The gluons correspond to the unstable gluon in our previous discussions\cite{cf,nielsen} with the use of
the time-like quantization. ( The unstable gluons in the time-like formulation have imaginary energies, while
the gluons corresponding to the unstable gluons can have real, but negative energies in the light cone formulation. )
Therefore, we obtain the following reduced Hamiltonian for analyzing the ground state of the system,
 
\begin{equation}
\label{hr}
H_r(\Phi)=-gB|\Phi|^2+\frac{g^2}{2}|\Phi|^4+\frac{g^2}{2}\rho_r\,\frac{1}{(-\partial_-^2)}\,\rho_r,
\end{equation}
with $\rho_r\equiv i(\Phi^{\dagger}\partial_-\Phi - \partial_-\Phi^{\dagger}\Phi) +\rho_{quark}$,
where we have put $\Phi\equiv \Phi_{p,\,n=0}$ for simplicity.
The field, $\Phi$, can be expressed by using creation and annihilation operators,

\begin{equation}
\label{p}
\Phi=\sqrt{\frac{\pi}{L}}\sum_{p>0,m=0,1,2,,}\frac{1}{\sqrt{2\pi p}}(a_{p,m}\phi_m(\vec{x})e^{-ipx}+
b_{p,m}^{\dagger}\phi_m^{\dagger}(\vec{x})e^{ipx}),
\end{equation}
where simplified notation such as $x=x^-$ and $p=p^+$ is exploited and
will be used below.
$\phi_m(\vec{x})=g_mz^m \exp(-|z|^2/4l^2)$ represents the normalized
eigenfunction of $\vec{D}^2$ with angular momentum, $m$, around the color magnetic field
in the lowest Landau level;
$\int d^2\vec{x} \phi_m^{\dagger}\phi_n=\delta_{m,n}$ with
$z=x_1+ix_2$, and
$g^2_m\equiv \frac{1}{\pi m!(2l^2)^{m+1}}$ with $l^2\equiv 1/gB$. 
$a_{p,m}$ and $b_{p,m}$ satisfy the commutation relations;
$[a_{p,m},a^{\dagger}_{k,n}]=\delta_{p,k}\delta_{m,n}, \quad [b_{p,m},b^{\dagger}_{k,n}]=\delta_{p,k}\delta_{m,n}$, and 
$\mbox{others}=0$.

\section{ground state}

When we express the first term in eq(\ref{hr}) in terms of the operators, $a_{p,m}$ and $b_{p,m}$,

\begin{equation}
\label{un}
\int_{-L}^{L}dx\,d^2\vec{x}:-gB|\Phi|^2:=-gB\sum_{p>0,m}\frac{1}{p}(a^{\dagger}_{p,m}a_{p,m}+b^{\dagger}_{p,m}b_{p,m}),
\end{equation}
we find that there exist states with lower energies than a trivial Fock vacuum, $|\mbox{vac}\rangle$; 
$a_{p,m}|\mbox{vac}\rangle =b_{p,m}|\mbox{vac}\rangle=0$. Namely, the gluons in the lowest Landau level
are produced spontaneously to form a state with lower energy than that of the vacuum.
( This fact that the Fock vacuum is not a real ground state
results from neglecting "zero modes" in the light cone quantization\cite{thorn,lcq}. )  
The production of the gluons is limited by the second term in eq(\ref{hr}) representing 
repulsion among the gluons. 
This is similar to the case of Higgs model. 
Actually, the Hamiltonian in eq(\ref{hr})
looks like the one of Higgs model.
But, there are several differences we should note between these two models.
The first one is that the field, $\Phi$ has no spatially uniform components; it is written
in terms of the wave functions, $\phi_m$, in the lowest Landau level.
On the other hand, Higgs field has a spatially uniform component with zero momentum. 
The second one 
is that there are infinitely many degenerate unperturbative states
in eq(\ref{un}), for instance, the state, $|p,m\rangle=a^{\dagger}_{p,m}|\mbox{vac}\rangle$,
is degenerate with the states, $|p,n\rangle$ ( $n\neq m$ ). 
On the other hand, there is no such degeneracy in Higgs model.
We will see below that the presence of
the infinitely many degenerate unperturbative states give rise to
the degeneracy in the ground state.

According to the standard procedure,
we minimize the classical energy of the field, 
in order to find the ground state of the Hamiltonian in eq(\ref{hr}).
First, we minimize the Coulomb
energy in eq(\ref{hr}). Since the energy is positive semi definite,
the field configuration giving zero Coulomb energy minimizes the energy.
Such a field is given by $\Phi_c=\exp(-ip_0x)\phi(\vec{x})$ with arbitrary $p_0=n_0\pi/L\neq 0$.
That is, the dependence of the longitudinal momentum
of the field is determined by minimizing the Coulomb energy.  
( We will discuss below more appropriate treatment of the Coulomb energy in the
ground state. )
 
Inserting the field, $\Phi_c=\exp(-ip_0x)\phi(\vec{x})$, into the remaining two terms in 
Hamiltonian, eq(\ref{hr}),
we find that

\begin{equation}
\label{hrr}
\frac{H_r(\Phi_c)}{2L}
=\int\frac{dx^{-}d\vec{x}}{2L}\Bigl(-gB|\Phi_c|^2+\frac{g^2}{2}|\Phi_c|^4 \Bigr)
=-gB\sum_n |a_n|^2+\frac{g^2}{2}\sum_{n_i} a_{n_1}^{\ast}a_{n_2}^{\ast}a_{n_3}a_{n_4}H_{n_1,n_2,n_3,n_4}
\end{equation}
where we have put

\begin{equation}
\Phi_c=\exp(-ip_0x)\sum_{n=0,1,2,,,} a_n\phi_n(\vec{x}), \quad \mbox{and} \quad 
H_{n_1,n_2,n_3,n_4}\equiv\int d\vec{x}\,\phi_{n_1}^{\ast}\phi_{n_2}^{\ast}\phi_{n_3}\phi_{n_4}.
\end{equation}
Hence, the configurations minimizing the energy of eq(\ref{hrr}) must satisfy

\begin{equation}
\label{T}
0=-gBa_n+g^2a_{n_2}^{\ast}a_{n_3}a_{n_4}H_{n,n_2,n_3,n_4}\equiv \sum_m T_{n,m}a_m
\end{equation}
with $T_{n,m}\equiv -gB\delta_{n,m}+g^2\int \,d\vec{x}|\Phi_c|^2\phi_n^{\ast}\phi_m $.
For non-trivial solutions of $a_n$ to exist, the determinant of $T$ must vanish; det$T=0$.

One of such solutions is given by 

\begin{equation}
\Phi_c=a\exp(-ip_0x)\phi_0(\vec{x}), \quad \mbox{with} \quad a=\frac{\sqrt{4\pi}}{g}.
\end{equation}
There are other solutions degenerate with this solution.
Namely, the classical ground state is not determined uniquely.
This is due to the presence of infinitely degenerate unperturbative states
as we have pointed above.

The color charge density, $\rho_r$, and the longitudinal momentum, $P_L$, of these solutions are given by

\begin{equation}
\rho_r=2p_0|\phi(\vec{x})|^2+\rho_{quark} \quad \mbox{and} \quad
P_L=2Lp_0^2\sum_m |a_m|^2, 
\end{equation}
The color charge, $Q=\int_{-L}^{L} dx^-\rho_r$ must vanish for the operator, $1/\partial_-^2$ to be well defined.
Furthermore, In order for $H_r(\Phi_c)P_L$ to be finite as $L\to \infty$, $p_0$ goes to zero 
in the limit,
just as $p_0\propto 1/L\to 0$. Note that the quantity of $H_r(\Phi_c)P_L$ is invariant\cite{cone}
under longitudinal boosts, just like masses of particles.

Here we wish to comment that as far as the classical solutions of the ground state are concerned,
the color charge, $Q$ does not vanish. 
Since $|\phi(\vec{x})|^2=\sum_{n,m}a_n^{\ast}a_m\phi_n^{\ast}(\vec{x})\phi_m(\vec{x})$ 
is never uniform in $\vec{x}$, the color charge of gluons does not cancel the color charge of quarks, $\rho_{quark}$,
which is assumed to be uniform.
The classical solutions correspond to coherent states. Thus, the fact that the classical color charge, $Q$
does not vanish, implies that the real ground state is not a coherent state of the field, $\Phi$.  

More appropriate treatment of the ground state is in the following.
Gluons are produced due to the term, $-gB|\Phi|^2$, in the Hamiltonian, but
they never condense to form the classical solutions as discussed just above. We speculate that
they form a state with approximately definite number of particles, or anti-particles
created by the operators, $a_{p_0,m}^{\dagger}$ or $b_{p_0,m}^{\dagger}$, that is,
the eigenstates of the number operators; $a_{p_0,m}^{\dagger}a_{p_0,m}|G\rangle=p_0a(p_0,m)|G\rangle$ and 
$b_{p_0,m}^{\dagger}b_{p_0,m}|G\rangle =p_0b(p_0,m)|G\rangle$. Here, we denote
the number of the particles, $a$, or
anti-particles, $b$ as
$p_0a(p_0,m)$ or $p_0b(p_0,m)$, respectively. Assuming that the ground state is composed of 
such eigenstates of the number operators, we evaluate\cite{06} the expectation value of the Hamiltonian,

\begin{eqnarray} 
\label{energy}
\langle G|:H_r:|G \rangle =-&gB&\sum_{p>0,m}\Bigl(a(p,m)+b(p,m)\Bigr)+ \nonumber \\
&+&\frac{g^2}{2L}\sum_{p,q>0,m,n}
\Bigl(a(p,m)a(q,n)+b(p,m)b(q,n)+2a(p,m)b(q,n)\Bigr)N_{m,n} \nonumber \\
&+&\frac{g^2}{2L}\sum_{p\neq q>0}\biggl(\frac{p+q}{p-q}\biggr)^2\sum_{m,n}
\Bigl(a(p,m)a(q,n)+b(p,m)b(q,n)\Bigr)N_{m,n}\nonumber \\
&+&\frac{g^2}{2L}\sum_{p,q>0}\biggl(\frac{p-q}{p+q}\biggr)^2\sum_{m,n}2a(p,m)b(q,n)N_{m,n},
\end{eqnarray}
with $H_{m,n}\equiv H_{m,n,m,n}=(m+n)!\,(\pi m!\,n!\,2^{m+n+2}\,l^2)^{-1}$,
where the Hamiltonian is normal-ordered.
The condition of $p\neq q$ in the third term comes from the regularization of the operator, $1/\partial_{-}^2$
in the Coulomb energy.

In eq(\ref{energy}) the first term represents the kinetic energy in the Landau level
and the second term does the energy of the repulsion between gluons.
These two terms are denoted by $E$.
The third and forth terms represent the Coulomb energy between gluons 
and are denoted by
$E_{Coulomb}$.
Obviously, these terms are non negative except for the first one.
Therefore, we can find a ground state with the lowest energy, $\langle G|:H:|G\rangle\equiv E+E_{Coulomb}$,
by minimizing the first two terms, $E$ and the last two terms, $E_{Coulomb}$,
respectively. 

It is easy to minimize the first two terms, i.e. $E$. The minimum of $E$ is given by a set of values,
$c^+(m)\equiv \sum_{p>0}(a(p,m)+b(p,m))$, 
since $E=-gB\sum_m c^+(m)+g^2/(2L)\sum_{m,n}c^+(m)H_{m,n}c^+(n)$.
Thus, minimizing the energy, $E$, does not
determine the distribution of the longitudinal momentum, $p=p^+$.
in $a(p,m)$ or $b(p,m)$. 
It simply gives the summation over the momentum, $p$, namely, $c^+(m)\propto gBL/g^2$. 
The dependence on $p$ is determined only by minimizing
the Coulomb energy, $E_{Coulomb}$. The energy, $E_{Coulomb}\geq 0$, can be minimized easily by assuming
that the ground state depends only on a single momentum, $p=p_0$, that is, $a(p,m)\propto \delta_{p,p_0}$
and $b(q,n)\propto \delta_{q,p_0}$. Namely, the ground state involves only gluons with the single 
longitudinal momentum, $p_0$.
This distribution of the momentum 
leads to the minimum, $E_{Coulomb}=0$. 
( Any other distributions with the dependence on various momenta 
give rise to higher energies ( $>0$ ). )
This fact is very result
used when we have discussed classical solutions, $\Phi_c\propto \exp(-ip_0x)$, of the ground state.

We should note that there are infinitely many solutions, $c^{+}(m)=a(p_0,m)+b(p_0,m)$, 
minimizing the energy, $E(c^+)+E_{Coulomb}$, in eq(\ref{energy}).
This is because $c^{+}(m)=a(p_0,m)+b(p_0,m)=a(p_0,m)-1/p_0+b(p_0,m)+1/p_0$.
Namely, the creation of an anti-particle with quantum number, $m$, 
along with the annihilation of a particle with quantum number, $m$,
does not change the energy, $E(c^{+})$ with $E_{Coulomb}=0$. 
This implies the presence of infinitely many excitations with zero energy
in a ground state $|G\rangle$.

Using the state, $|G\rangle$, the color charge is 
given by $Q=\int_{-L}^{L} dx^- \langle G| \rho_r|G \rangle=2p_0\sum_m \Bigl(a(p_0,m)-b(p_0,m)\Bigr)|\phi_m(\vec{x})|^2+2L\rho_{quark}$.
 In order for this charge to be uniform, 
it is sufficient to take $\Bigl(a(p_0,m)-b(p_0,m)\Bigr)$ as be independent of $m$,
since $\sum_m |\phi_m(\vec{x})|^2=gB/2\pi$.
Putting 
$c^-(p_0)\equiv\Bigl(a(p_0,m)-b(p_0,m)\Bigr)$, 
we can rewrite the color charge such as
$Q=p_0 c^-(p_0)gB/\pi+2L\rho_{quark}$. Therefore, the color charge can vanish when we take $p_0c^-(p_0)=-2\pi L\rho_{quark}/gB$.
The assumption that the ground state is an approximate eigenstate of the number operators
is consistent with the condition of the color charge, $Q$, to vanish.

The longitudinal momentum also vanishes in the limit of $L\to \infty$.
The longitudinal momentum is given such as
$P_L=p_0\sum_m (p_0a(p_0,m)+p_0b(p_0,m))=p_0^2\sum_m c^{+}(m)$. That is, it is the sum of 
momentum, $p_0$, each particles carries. 
We remind you that the number of the particles with $p=p_0$ and angular momentum, $m$, is 
given by $p_0a(p_0,m)+p_0b(p_0,m)$.
Since $c^{+}(m)$ ( $p_0$ ) goes to infinity ( zero ) such as $c^{+}\propto L$ ( $p_0\propto 1/L$ ) 
in the limit of $L\to \infty$, 
$P_L$ goes to zero.

In the above discussion about classical solutions
we have assumed that the ground state is a coherent state of the field operator, $\Phi$.
But, this leads to the nonvanishing color charge although it must vanish for the regularity of $1/\partial_{-}^2$.
In the more appropriate argument, we have assumed that the ground state is 
an eigenstate of the number operators. 
This leads to the color charge to vanish.
When the number, $p_0a(p_0,m)+p_0b(p_0,m)$, of the particles in each state specified by $m$ is sufficiently large,
we may use approximately the coherent state for such a state. Then,  
the ground state can be represented by the classical solutions, although $\langle G|\Phi|G \rangle=0$.
In the case
classical Coulomb energy should be taken to vanish for consistency.
In the next section, we examine excitation energies by using the classical ground state.

\section{excitation energies on the ground state, $\langle \Phi \rangle=\Phi_c$} 

Now, we wish to examine the excitation energies and show\cite{06} that the gluons
form effectively a two dimensional quantum well.
Supposing that the classical solutions, $\Phi_c$,
approximately describe the ground state, we put the field operator
such as $\Phi=\Phi_c+\delta\Phi$ in the Hamiltonian, $:H_r:=H_0+H_{Coulomb}$, and
diagonalizing it by taking only quadratic terms of $\delta\Phi$, 

\begin{eqnarray}
\label{h0}
H_0=\int^{L}_{-L}dx(-gB|\Phi_c+\delta\Phi|^2+\frac{g^2}{2}|\Phi_c+\delta\Phi|^4)\simeq 
\int^{L}_{-L}dx(-gB|\Phi_c|^2+\frac{g^2}{2}|\Phi_c|^4) \nonumber \\
-\sum_{p>0}\frac{gB}{p}(\alpha_p^{\dagger}\alpha_p+\beta_p^{\dagger}\beta_p)
+\sum_{p>0}\frac{2g^2|\Phi_c|^2}{p}(\alpha_p^{\dagger}\alpha_p+\beta_p^{\dagger}\beta_p)
+\sum_{p>0}\frac{g^2}{p}(\alpha_p^{\dagger}\beta_pe^{i2p_0x}\Phi_c^2+h.c.)
\end{eqnarray}
and 
\begin{eqnarray}
\label{hc}
H_{Coulomb}&=&\frac{g^2}{2}\int^{L}_{-L}dx \rho_r\frac{1}{-\partial_{-}^2}\rho_r\simeq
g^2\sum_{0<p\neq p_0}\frac{1}{p}\Bigl(\frac{p+p_0}{p-p_0}\Bigr)^2\Bigl(\alpha_p^{\dagger}\alpha_p+\beta_p^{\dagger}\beta_p\Bigr)
|\Phi_c|^2+ \nonumber \\ 
&+&\frac{g^2}{2}\sum_{0<p,q \neq p_0}\frac{1}{\sqrt{pq}}\frac{(p+p_0)(q+p_0)}{(p-q)^2}\delta_{2p_0,p+q}
\Bigl(\alpha_p\alpha_q e^{-2ip_0x}\Phi_c^{\ast\,2}+h.c.\Bigr)- \nonumber \\
&-&g^2\sum_{0<p}\frac{(p+3p_0)(p-p_0)}{\sqrt{p(p+2p_0)}(p+p_0)^2}\Bigl(\alpha_{2p_0+p}^{\dagger}\beta_p e^{2ip_0x}\Phi_c^2+h.c. \Bigr),
\end{eqnarray}
where

\begin{equation}
\delta\Phi=\sqrt{\frac{\pi}{L}}\sum_{0<p}\frac{1}{\sqrt{2\pi p}}(\alpha_p(\vec{x})e^{-ipx}+\beta^{\dagger}_p(\vec{x})e^{ipx})
\end{equation}
with $\alpha_p(\vec{x})\equiv\sum_{m=0,1,2,,}a_{p,m}\phi_m(\vec{x})$ and $\beta_p(\vec{x})\equiv\sum_{m=0,1,2,,}b_{p,m}\phi_m(\vec{x})$. 
We have assumed that the Coulomb energy of the classical solutions, $\Phi_c$, vanishes.

It is interesting to see that the modes of $\alpha_p$ and $\beta_p$ with different longitudinal momenta do not mix with each other
in $H_0$ and in the first term of $H_{Coulomb}$. The mixing between the modes with
different momenta only arises
owing to the interactions represented by the remaining two terms in $H_{Coulomb}$. 
Furthermore, the modes with $p=p_0$ does not mix with the other modes with $p\neq p_0$ even
if we take into account all of the interactions in eq(\ref{h0}) and eq(\ref{hc}). This implies that the "transverse modes" with $p=p_0$ decouples from
the other "longitudinal modes" with $p\neq p_0$. 

It is easy to diagonalize the transverse modes,

\begin{eqnarray}
\label{d}
\int d\vec{x}\,\frac{1}{p_0}\Biggl((-gB+2g^2|\Phi_c|^2)(\alpha^{\dagger}_{p_0}\alpha_{p_0}+\beta^{\dagger}_{p_0}\beta_{p_0})
+g^2(e^{2ip_0x}\Phi_c^2\alpha_{p_0}\beta_{p_0}+h.c.)\Biggr) \nonumber \\
=\int d\vec{x}\,\frac{1}{p_0}\Biggl((-gB+3g^2|\Phi_c|^2)\alpha'^{\dagger}_{p_0}\alpha'_{p_0}
+(-gB+g^2|\Phi_c|^2)\beta'^{\dagger}_{p_0}\beta'_{p_0}\Biggr) 
\end{eqnarray}
with $\alpha'_{p_0}\equiv (\alpha_{p_0}+\beta_{p_0})/\sqrt{2}$
and $\beta'_{p_0}\equiv (\alpha_{p_0}-\beta_{p_0})/\sqrt{2}$.
Here, the second term can be rewritten as

\begin{equation}
\int d\vec{x}\,\frac{1}{p_0}(-gB+g^2|\Phi_c|^2)\beta'^{\dagger}_{p_0}\beta'_{p_0}=\frac{1}{p_0}\sum_{n,m}T_{n,m} b'^{\ast}_nb'_m
\end{equation}
where $T_{n,m}$ has been defined in eq(\ref{T}) and
$\beta'_{p_0}$ has been expanded in term of the eigenfunctions of the lowest Landau level;
$\beta'_{p_0}\equiv \sum_mb'_m\phi_m(\vec{x})$. Since det$T=0$ as explained in eq(\ref{T}),
the transverse modes are gapless.
On the other hand, as the first term in eq(\ref{d}) can be rewritten as
$(-gB+3g^2|\Phi_c|^2)=3(-gB+g^2|\Phi_c|^2)+2gB$, the eigenvalues of 
the operator, $\int d\vec{x}\,(-gB+3g^2|\Phi_c|^2)$,
are positive definite. Thus, the corresponding modes are gapped.

Therefore, the transverse modes of the excitations on the ground state, $\langle \Phi\rangle=\Phi_c$,
are gapless. This is due to the presence of infinitely degenerate unperturbative states
in the lowest Landau level. The degeneracy is not lifted up by the repulsive 
self-interaction,
or the Coulomb interaction.
 
Here we should make a comment.
Rigorously speaking, we have simply shown the existence of zero eigen value of $T$, but have not yet
shown that the eigenvalues are positive semi definite. 
The condition of the semipositivity must hold since the Hamiltonian
is bounded below. At least, we can choose appropriate solutions, $\Phi_c$,
for the eigenvalues of $T(\Phi_c)$ to be positive semi definite.
In the discussion we have assumed that we take such classical solutions.

We now proceed to diagonalize the Hamiltonian in
the longitudinal components, $\alpha_{p\neq p_0}$ and $\beta_{p\neq p_0}$.
Before analyzing the problem, we should note that the reference momentum, $p_0$,
characterizing the ground state must go to zero as $L\to \infty$.
Thus, we consider only such modes with components of $p\gg p_0$.
Then, it is easy to extract a Hamiltonian involving the longitudinal modes from $H_0+H_{Coulomb}$,

\begin{equation}
\int d\vec{x}\,(-gB+3g^2|\Phi_c|^2)\sum_{p\gg p_0}\frac{1}{p}(\alpha^{\dagger}_p\alpha_p+\beta^{\dagger}_p\beta_p).
\end{equation} 

As we have made a comment just above, the eigenvalues	of 
the operator, $\int d\vec{x}\,(-gB+3g^2|\Phi_c|^2)$,
are positive definite.  	
Thus, we find that the longitudinal modes have gap energies on the ground state, $\langle \Phi\rangle=\Phi_c$.

It is important to note that the gaps of the longitudinal modes
arise due to the Coulomb interaction. Actually, if we neglect the interaction,
a Hamiltonian of the longitudinal modes is the same as the one of the transverse modes,
which is given in eq(\ref{d}). Thus, the gaps do not arise.

We have discussed the cases of the modes with $p=p_0$ and $p\gg p_0$.
Here, we should make a comment the case of modes with $p\sim p_0$, but $p\neq p_0$.
In this case the modes couple with other modes with different longitudinal momentum
so that the diagonalization of the relevant Hamiltonian is very difficult.
But we can diagonalize the Hamiltonian only by taking the modes with $p_{\pm}=p_0\pm \pi/L$
and neglecting the other modes. These modes receive large Coulomb energies due to the terms,
$\sim 1/(p_{\pm}-p_0)^2$. Then, we find 
that there is no gapless mode. This is 
not exact treatment, but a simple exercise to grip real excitation energies.
It makes us speculate that
the energy gaps exist even if we diagonalize the Hamiltonian
including all modes with $p\sim p_0$. 

Up to now, we have examined excitation energies by using the classical ground state, $\Phi_c$.
On the other hand, by using the quantum ground state $|G\rangle$,
we have given a
plausible argument in the previous paper\cite{lcq} that there arise the gap energies in
the longitudinal direction.
In the argument the Coulomb energy also plays an important role for the 
generation of the gap.

The fact that there exist the gapless modes in the transverse directions, 
while no gapless modes in the longitudinal direction implies that
gluons are localized in a two dimensional layer.
They can move easily in the transverse plane, but
can not move in the longitudinal direction as far as we are concerned with
sufficiently low energies.
This is very similar to the case of electrons confined 
in a two dimensional quantum well
fabricated in semiconductors.
Electrons can move only within the two dimensional quantum well as far as
we are concerned with low energies or low temperatures.

\section{zero modes and quantum Hall states of gluons}

We have shown that there are zero modes in the transverse directions,
in other words, the ground state, $\langle \Phi \rangle\neq 0$, is degenerate with those states involving
the excitations of the zero modes. 
As we can see soon below, there are infinitely many zero modes.
It suggests that the ground state discussed above is unstable.
In general,
the excitations of the zero modes lead to a unique stable ground state owing to the residual interactions, 
which have been
neglected in the above approximation.   
Thus, we need to answer the question what is the real stable ground state.
We will find that the real ground state is a quantum Hall state of the zero mode gluons\cite{cf,qhs}.

In order to do so, by taking a simple classical ground state, $\Phi_c$, we first show explicitly 
that there are infinitely many zero modes. 
After that,
we examine the effects of the zero mode excitations.
( We have already shown that there are infinitely many zero modes by using the quantum ground state $|G\rangle$.
In this section we analyze classical zero mode solutions around the classical ground state. )

The zero modes are given by the solutions of the following equations,

\begin{equation}
\label{zero}
\sum_{m=0,1,2,,,}T_{n,m}a_m=\sum_{m=0,1,2,,,}\Bigl(-gB\delta_{n,m}+g^2\int d\vec{x}|\Phi_c|^2\phi^{\ast}_n\phi_m\Bigr)a_m=0,
\end{equation}
with $\delta\Phi_0=\sum_{m=0,1,2,,,}a_m\phi_m$,
where $\Phi_c$ represents a classical solution of a ground state, which also satisfies
the same equation as the equation (\ref{zero}).
As an example we consider a field configuration such as $\Phi=a_0\phi_0$.
Obviously, it is a solution in eq(\ref{zero}) with $a_0=\sqrt{4\pi/g^2}$.
It is not, however, a solution representing a ground state,
because we expect that the ground state should be spatially uniform.
Although we have mentioned the absence of such a spatially uniform
solutions in the lowest Landau level,
we may consider an approximate solution
which is spatially uniform, that is, $\Phi_{const}=\mbox{const.}\times e^{-ip_0x}$.
Using the approximate classical ground state,
we solve the equation for the zero modes,

\begin{equation}
\sum_mT_{n,m}a_m=\sum_m(-gB+g^2|\Phi_{const}|^2)\delta_{n,m}a_m=(-gB+g^2|\Phi_{const}|^2)a_n=0.
\end{equation} 
 Hence, we find that any configurations of $\delta\Phi_0$ satisfy the equation if we take $|\Phi_{const}|=gB/g^2$.
The configuration, $|\Phi_{const}|=gB/g^2$, is just the field configuration
minimizing the potential energy, $-gB|\Phi|^2+g^2|\Phi|^4/2$ when we neglect the limitation of the field
such as it occupies the lowest Landau level.
Therefore, 
any modes in the lowest Landau level can be zero modes.
Obviously, there are infinitely many independent zero modes.

We proceed to show that the excitations of these zero modes
partially screen the color magnetic field, $B$. The result is expected naively from
the fact that the ground state is a condensed state of the color charged field, $\Phi$,
just like a color superconductor. Thus, it apparently seems that
the color magnetic field is ejected or squeezed in the condensed state
of the field, $\Phi$. But this is not true as we will show below.

In order to see the partial screening, we calculate color magnetic field generated by
the zero modes. As an explicit example, we take a zero mode described by the wave function, $\phi_m=g_mz^m\exp(-|z|^2gB/4)$.  
The mode generates a color current in the transverse directions given by

\begin{equation}
\delta \vec{J}=ig\langle \Bigr(\delta\Phi^{\dagger}_0 \vec{D}\delta\Phi_0-(\vec{D}\delta\Phi_0)^{\dagger}\delta\Phi_0\Bigr)\rangle
\quad \mbox{with} \quad \vec{D}\equiv\vec{\partial}+ig\vec{A^{B}}
\end{equation} 
where the expectation value is calculated by using an eigenstate of the number operator of the zero mode.
The zero mode, $\delta\Phi_0$, is given by

\begin{equation}
\delta\Phi_0=\Bigl(\frac{a_{p_0,m}\phi_m e^{-ip_0x}+b^{\dagger}_{p_0,m}\phi^{\ast}_m e^{ip_0x}}{\sqrt{2\pi Lp_0}}\Bigr)_{zero}
=\frac{b'_{m}\phi_m e^{-ip_0x}-b'^{\dagger}_{m}\phi^{\ast}_m e^{ip_0x}}{2\sqrt{\pi Lp_0}}
\end{equation}
with an operator of non zero mode, $a'_{m}=(a_{p_0,m}+b_{p_0,m})/\sqrt{2}$, and 
the operator of the zero mode, $b'_{m}=(a_{p_0,m}-b_{p_0,m})/\sqrt{2}$.
This color current induces a color magnetic field, $\delta B$,
assumed to point to the longitudinal
direction. The field satisfies 
the Maxwell equation,

\begin{equation}
J_z\equiv J_1+iJ_2=ig\langle :\Bigl(\delta\Phi^{\dagger}_0D_z\delta\Phi_0-(D_{\bar{z}}\delta\Phi_0)^{\dagger}\delta\Phi_0\Bigr):\rangle
=-i\partial_z\delta B,
\end{equation}
where we have assumed the rotational symmetry of the induced magnetic field around the longitudinal direction
and have used the following notations, 

\begin{eqnarray}
D_{z}\equiv \partial_1+i\partial_2+ig(A^{B}_1+iA^{B}_2)=\partial_z-gBz/2 \nonumber \\ 
D_{\bar{z}}\equiv \partial_1-i\partial_2+ig(A^{B}_1-iA^{B}_2)=\partial_{\bar{z}}+gB\bar{z}/2, 
\end{eqnarray}
with $z\equiv x_1+ix_2$ and a gauge $A^{B}_1=-Bx_2/2$ and $A^{B}_2=Bx_1/2$. 
It is easy to evaluate the current of $J_z$,

\begin{equation}
J_z=-i\frac{gB\, g^2_m z|z|^{2m}}{2\pi Lp_0}\exp(-\frac{gB|z|^2}{2})\,N_m,
\end{equation}
with the number of the zero mode, $N_m \equiv \langle b'^{\dagger}_m b'_m \rangle$ in the state.
Thus, the color magnetic field induced by the current is 

\begin{equation}
B(|\vec{x}|)=\delta B+B=-\frac{g^2B g^2_m}{4\pi Lp_0}(\frac{2}{gB})^{m+1}\Gamma(1+m, \frac{gB|\vec{x}|^2}{2})N_m
+B
\end{equation}
where $\Gamma(a,z)$ is the incomplete gamma function; $\Gamma(a,z)=\int^{\infty}_{z}t^{a-1}e^{-t}dt$.
The value of $Lp_0$ is finite as $L\to\infty$.
The first term represents the induced magnetic field, $\delta B$, and the second one does the original magnetic field, $B$.
Therefore, we find that the excitations of the zero modes partially screen the color magnetic field.

It seems apparently that the magnetic field is completely screened
by the excitations of the zero modes just as in superconductors.
This is owing to the existence of the condensed color charged scalar field, $\Phi$,
in the ground state of the ferromagnetic quark matter.
But, it is not so analogous.
In the superconductors the existence of magnetic field
is energetically unfavorable; the magnetic field destroys Cooper pairs of electrons.
This leads to expelling the magnetic field or squeezing the magnetic flux to form
vortices. 
But in the case of QCD the existence of the color magnetic field is energetically favorable 
because it decreases the energy of effective potential;
the potential has a nontrivial minimum at non zero color magnetic field.
The condensed state of $\Phi$ arises owing to the presence of the color magnetic field.
Hence, the completely screening of the color magnetic field does not arise.

As we have shown, there are infinitely degenerate states in the ground state, $\langle \Phi\rangle =\Phi_c$
since there exist infinitely many zero modes. 
The ground state with such a degeneracy may be unstable against residual interactions between the zero modes 
which we have neglected. Stable states must be gapped. That is,
there are no excitations with zero energy.
A candidate for the state is a quantum Hall state of the zero mode gluons.
A finite gap exists in the state even if the quantum Hall state is composed of
bosons like gluons\cite{nakajima}. It is well known\cite{qhs} 
that 
a quantum Hall state of electrons arises due to the repulsive interaction between the electrons just like
Coulomb interaction,
even if the interaction is fairly weak. 
The electrons in two dimensional quantum well
occupy the lowest Landau level under external magnetic field
and have infinitely many degenerate states unless we take into account
repulsive Coulomb interaction.
The same thing holds even if gluons ( bosons ) are present
in the two dimensional well. 
Actually, there is repulsive self-interaction, $|\delta \Phi_0|^4$, between the zero modes.
The interaction leads to the formation of the quantum Hall state of the zero mode gluons.
In order for the state to appear,
the two dimensional color charge density of
the gluons, $\langle \rho_r\rangle l_d$, must satisfy
the conditions on the filling factors, $\frac{2\pi\langle \rho_r\rangle}{gBl_d}$,

\begin{equation}
\label{filling}
\frac{2\pi\langle \rho_r\rangle l_d}{gB}=
\frac{2\pi\langle i(\delta\Phi^{\dagger}_0\partial_-\delta\Phi_0 
- \partial_-\delta\Phi^{\dagger}_0\delta\Phi_0) \rangle l_d}{gB}=
\frac{1}{2n}=1/2,\,\,1/4\,\,,\,,\,,
\end{equation} 
with integer, $n$,
where $l_d$ denotes a width of the quantum well. 
The width is given approximately by the inverse of the gap energy in the longitudinal direction.
The appearance of the even number in the denominator is due to
the fact that gluons are bosons. It is odd number if gluons 
were fermions. Such a color charge density may arise automatically by generating the zero modes
for the quantum Hall state
to be formed.
Therefore, the quantum Hall state of the gluons naturally arises owing to the
the excitations of the zero mode gluons in the two dimensional quantum well.
Please refer our previous papers\cite{cf} for the more detail treatment of the quantum Hall states by the use
of Chern-Simons gauge theory.

\section{summary and discussion}

Using the light cone quantization with neglecting "zero modes",
we have shown that gluons in the ferromagnetic quark matters
effectively form a two dimensional quantum well.
Namely, the gluons with a longitudinal momentum, $p^+_0$, characterizing the ground state,
decouple with the other gluons with longitudinal momenta, $p^+\neq p^+_0$. 
The gluons with $p^+=p^+_0$ are gapless; their excitation energies 
start to grow at zero energy. On the other hand, the gluons with $p^+ \gg p^+_0$
which also do not couple with other gluons with different momentum, $q^+\neq p^+$,
are gapped; their excitation energies start to grow at a finite non zero energy.
The gluons with $p^+\sim p^+_0$ couple with the other gluons with 
different momentum, $q^+\neq p^+$ ( $q^+\sim p^+_0$ ). Although 
we have not discussed in detail the case of the gluons with the momentum, $p^+\sim p^+_0$,
a simple exercise shows that such gluons also possess gap energies.
This feature of the excitation spectra implies that 
the gluons form the two dimensional
quantum well. 

The formation of the two dimensional quantum well in the ferromagnetic quark matter is 
a necessary condition for the realization of a quantum Hall state of gluons.
The state can be realized in such a quantum well as in the case of semiconductors.
We have discussed that since there are infinitely many excited states with zero energy ( zero modes ),
they are produced without any energy costs to form a quantum Hall state
owing to the repulsive interaction between them.
The quantum Hall state has an energy gap, i.e. no zero modes.
Thus, it is the stable ground state of the gluons
in the quark matter.

Such a quark matter may be present in neutron stars.
We may suppose that the quantum layers parallel to each other are 
formed in the quark matter of the neutron stars.
That is, the quark matter has a domain structure
in which many parallel layers are involved and the color magnetic field points to a direction
in the domain. Then, 
there is a quantum Hall state of gluons
in each layer. The state possesses the color charge of the gluons.
The sign of the charge is common in all layers.
The amount of the color charge in a layer
can be determined by the condition eq(\ref{filling}) on the filling factor.
It must be compensated by the color charge of quarks due to 
the condition of the color neutrality. Therefore, the gas of the quarks carries the color charge
and rotates around the color magnetic field.
Since the gas of quarks also carries electric charges, its rotation 
induces an electromagnetical magnetic field. ( On the other hand,
electrons whose electric charges compensate the charge of quarks, do not rotate around the color magnetic field so that
they do not produce an electromagnetical magnetic field. )
Thus, the electromagnetical magnetic field is generated spontaneously in the domain.
The domain may be extended all over the quark matter in neutron stars.  
This is a possible origin\cite{cf2,magnetar} of the strong magnetic field
observed in magnetars.

Recent analyses\cite{gapless} of gapless color superconductors indicate the existence of the phase transition from
the color superconducting state to the color ferromagnetic state with decreasing the baryon density.  
Actually, it has been shown that a color magnetic instability in the color superconductivity
arises, that is, the instability such that external color magnetic field can penetrate
the color superconductor.
In other words, the coefficient of the kinetic term, $|\vec{D}\Phi_d|^2$ of Higgs fields, $\Phi_d$, representing
diquark pairs becomes negative. 
Furthermore, it has been shown  
that the spontaneous generation of color magentic field ( spatially inhomogeneous gauge fields )
arises. 
These results have been obtained only by the analyses of quark 
dynamics. But, it strongly suggests that we need to include appropriately the quantum effects of
the gluons along with those of the quarks, just as we have done in the present paper.

We also wish to point out an intriguing relationship. 
That is, a relation between color glass condensate of gluons in nucleons
and quantum Hall state of gluons in quark matter.
It is quite plausible that
the color ferromagnetic quark matter arises through the phase transition
from hadronic phase when we increase the baryon density of nuclear matter.
Then, it is natural to ask how a gluonic specific state of 
color glass condensate in nucleons
transforms into the gluonic state mentioned above, namely, 
the quantum Hall state of gluons. Since the saturation momentum of 
the color glass condensate increases with baryon density, 
the gluons with even large $x=p^{+}/P_L\sim 1$, may constitute
the color glass condensate in the dense quark matter. On this point, we have argued\cite{cgciwa} a similarity between
the color glass condensate and the quantum Hall state of gluons. 
We wish to clarify the relation much more in near future.

\vspace*{2em}
This work was supported by Grants-in-Aid of the Japanese Ministry
of Education, Science, Sports, Culture and Technology (No. 13135218).



\begin{thebibliography}{99}
\bibitem{cs}B.C. Barrois, Nucl.Phys. B129 (1977) 390. \\
D. Bailin and A. Love, Phys. Rep. 107 (1984) 325. \\
K. Rajagopal and F. Wilczek, hep-ph/0011333.
\bibitem{cf}A. Iwazaki and O. Morimatsu, Phys. Lett. B 571 (2003) 61.\\ 
A. Iwazaki, O. Morimatsu, T. Nishikawa and M. Ohtani,
Phys. Lett. B 579 (2004) 347; Prog. Theor. Phys. Suppl. 156 (2004) 178;
Phys. Rev. D 71 (2005) 034014.
\bibitem{tatsumi}T. Tatsumi, T. Maruyama and E. Nakano, Prog. Theor. Phys. Suppl. 153 (2004) 190.
\bibitem{savvidy}G.K. Savvidy, Phys. Lett. B 71 (1977) 133.
\bibitem{cf2}A. Iwazaki, O. Morimatsu, T. Nishikawa and M. Ohtani, Int.J.Mod.Phys. A22 (2007) 721.
\bibitem{gapless}M. Huang and I. A. Shovkovy, Phys. Rev. D 70, (2004) 051501,
K. Iida and K. Fukushima, hep-ph/0603179,
E. V. Gorbar, M. Hashimoto, V. A. Miransky and I. A. Shovkovy, Phys. Rev. D 73, (2006) 111502(R).
\bibitem{n-star}J. Lattimer and M. Prakash, astro-ph/0612440.
\bibitem{magnetar}A. Iwazaki, Phys. Rev. D72 (2005) 114003.
\bibitem{nielsen}J. Ambj\o{}rn, N.K. Nielsen and P. Olesen, Nucl. Phys. B152 (1979) 75.
\bibitem{ninomiya}H.B. Nielsen and M. Ninomiya, Nucl. Phys. B156 (1979) 1.\\
H. B. Nielsen and P. Olesen, Nucl. Phys. B160 (1979) 330.
\bibitem{06}A. Iwazaki, Phys. Rev. D75 (2007) 034020.
\bibitem{qhs} The Quantum Hall Effect, 2nd Ed. edited by R.E. Prange
  and S.M. Girvan ( Springer-Verlag, New York, 1990 ).
\bibitem{lc}P.A.M. Dirac, Rev. Mod. Phys. 21 (1949) 392.
\bibitem{cone}for a review, see A. Harindranath, hep-ph/9612244, 
T. Heintl, hep-th/0008096. \\ R. Venugopalan, nucl-th/9808023.
\bibitem{yamawaki}T. Maskawa and K. Yamawaki, Prog. Theor. Phys. 56 (1976) 270.
\bibitem{yama}S. Uehara, S. Yamada and K. Yamawaki, "Nagoya 2004, Dynamical symmetry breaking" (2004) 157.
\bibitem{thorn}J.S. Rozowsky and C.B. Thorn, Phys. Rev. Lett. 85 (2000) 1614.\\
V.T. Kim, G.B. Pivovarov and J.P. Vary, Phys. Rev. D69 (2004) 085008.\\
D. Chakrabarti,
A. Harindranath, L. Martinovic and J.P. Vary, Phys. Lett. B582 (2004) 196.\\
D. Chakrabarti, A. Harindranath
and J.P. Vary, Phys. Rev. D 71 (2005) 125012. 
\bibitem{lcq}A. Iwazaki, hep-ph/0702255, to be published in Phys. Rev. D.
\bibitem{cgc}E. Iancu, A. Leonidov and L. McLerran, hep-ph/0202270.\\
E. Iancu and R. Venugopalan, hep-ph/0303204.
\bibitem{nakajima} T. Nakajima and M. Ueda, Phys. Rev. Lett. 91 (2003) 140401.
\bibitem{cgciwa}A. Iwazaki, hep-ph/0604222.
\end{thebibliography}
\end{document}